\documentclass[preprint2]{aastex}

\usepackage{epsfig,graphicx}
\usepackage{subfigure}
\usepackage{amsmath}
\usepackage{amssymb}

\shorttitle{Astro2010 GRB White Paper}
\shortauthors{Stamatikos et al.}

\begin{document}

\title{Multi-Messenger Astronomy with GRBs: \\ A White Paper for the Astro2010 Decadal Survey}

\author{M. Stamatikos\altaffilmark{1}, N. Gehrels\altaffilmark{2}, F. Halzen\altaffilmark{3}, P. M$\acute{\text{e}}$sz$\acute{\text{a}}$ros\altaffilmark{4}, and P. W. A. Roming\altaffilmark{5}}
\altaffiltext{1}{\scriptsize{Center for Cosmology and Astro-Particle Physics Fellow, Department of Physics, Department of Astronomy, The Ohio State University, Columbus, OH 43210, USA. Correspondence to Michael.Stamatikos-1@nasa.gov (1-614-292-0734).}}
\altaffiltext{2}{\scriptsize{Astro-particle Physics Laboratory, Code 661, NASA/Goddard Space Flight Center, Greenbelt, MD 20771, USA}}
\altaffiltext{3}{\scriptsize{Department of Physics, University of Wisconsin, Madison, WI 53706, USA}}
\altaffiltext{4}{\scriptsize{Center for Particle Astrophysics, Department of Astronomy and Astrophysics, Department of Physics, Pennsylvania State University, University Park, PA 16802, USA}}
\altaffiltext{5}{\scriptsize{Department of Astronomy and Astrophysics, Pennsylvania State University, University Park, PA 16802, USA}}

\begin{abstract}
Gamma-ray Bursts (GRBs) are relativistic cosmological beacons of transient high energy radiation whose afterglows span the electromagnetic spectrum. Theoretical expectations of correlated neutrino and/or gravitational wave (GW) emission position GRBs at an astrophysical nexus for a metamorphosis in our understanding of the Cosmos. This new dawn in the era of experimental (particle) astrophysics and cosmology is afforded by current and planned facilities enabling the novel astronomies of high energy neutrinos and gravitational waves, in concert with unprecedented electromagnetic coverage. In this white paper, we motivate GRBs as a compelling scientific theme and highlight key technical advances that may facilitate fundamental breakthroughs in the context of Swift, Fermi, IceCube and LIGO (Laser Interferometer Gravitational Wave Observatory), whose capabilities would be augmented with JANUS (Joint Astrophysics Nascent Universe Satellite), EXIST (Energetic X-ray Imaging Survey Telescope) and LISA (Laser Interferometer Space Antenna). Scientific synergy will be achieved by leveraging the combined sensitivity of contemporaneous ground-based and satellite observatories, thus optimizing their collective discovery potential for: (i) revealing the origin(s) and acceleration mechanism(s) of cosmic rays, (ii) exposing GRB progenitor(s) and (iii) exploring the high-z Cosmos. Hence, the advent of GRB multi-messenger astronomy may cement an explicit connection to fundamental physics, via nascent cosmic windows, throughout the next decade and beyond.
\end{abstract}


\clearpage

\section{Introduction: A New Dawn in Multi-Messenger Astronomy for GRBs}
\label{Intro}

This science white paper, intended for the Stars and Stellar Evolution (SSE) Science Frontier Panel of the Astro2010 Decadal Survey, identifies GRBs as a compelling science theme with promising discovery potential towards answering fundamental questions regarding cosmology and particle astrophysics within the next decade, via: \emph{\textbf{(i) revealing the origin(s) and acceleration mechanism(s) of cosmic rays, (ii) exposing GRB progenitor(s) and (iii) exploring the high-z Cosmos.}} This new opportunity will be afforded by the technical means of performing multi-messenger observational campaigns owing to key advances in existing and future satellite- and ground-based facilities such as: Swift \citep{Gehrels:2007}, Fermi \citep{Atwood:2009}, IceCube \citep{Ahrens:2004}, LIGO (Laser Interferometer Gravitational Wave Observatory) \citep{Abramovici:1992}, JANUS (Joint Astrophysics Nascent Universe Satellite) \citep{Roming:2008}, EXIST (Energetic X-ray Imaging Survey Telescope) \citep{Grindlay:2009} and LISA (Laser Interferometer Space Antenna) \citep{Araujo:2007}. In addition, a fleet of international facilities are poised for correlated follow-up via a real-time alert system known as the GRB Coordinate Network (GCN) \citep{Barthelmy:2008}.

\section{The Limitations of (sub-MeV) GRB Electromagnetic Observations}

Thus far, a canonical phenomenological description, known as the fireball model \citep{Meszaros:1993b}, has successfully described most prompt and afterglow GRB electromagnetic observations. In this scheme, the generic mechanism responsible for generating their large energy ($\sim 10^{50} - 10^{52}$ ergs) and super-Eddington luminosity, within $\sim 0.01$ to $\sim1000$ sec, is the dissipation (via internal shocks, magnetic dissipation and external shocks) of highly relativistic kinetic energy, acquired by electrons and positrons Fermi accelerated in an optically thick, relativistically expanding plasma. The acceleration of electrons, in the intense magnetic field of the fireball, leads to the emission of prompt non-thermal photons, via synchrotron radiation and possibly inverse Compton scattering. The prompt electromagnetic emission\footnote{Energies as low as of order keV have been observed in X-Ray Rich (XRRs) GRBs and X-Ray Flashers (XRFs) \citep{Heise:2001}, while the Energetic Gamma Ray Experiment Telescope (EGRET) and Fermi \citep{Tajima:2008} have observed $\gamma$-rays with energy in excess of 100 MeV and up to GeV. TeV emission has been constrained \citep{Atkins:2005}, with the exception of a $\sim3\sigma$ event \citep{McEnery:1999}.} ranges from $\sim$ 10 keV $\lesssim\epsilon_{\gamma}\lesssim\sim$ 20 GeV with a characteristic photon (peak) energy in the $\gamma$-ray regime ($\overline{\text{E}}_{\text{peak}}\sim250$ keV), as observed by satellites such as the Compton Gamma-Ray Observatory (CGRO), Beppo-SAX, Swift and Fermi. External shocks, interacting with either the interstellar medium or the outer stellar envelope, result in multi-wavelength afterglow emission that spans the electromagnetic spectrum on time scales up to a year and are typically followed-up via the GCN.

Despite great strides made by satellite-and ground-based electromagnetic observations, the details of GRB progenitor(s) remain concealed. This physical limitation is due to the initial optical thickness of the fireball, prior to adiabatic relativistic expansion, which precludes us from witnessing the genesis of a GRB. The relativistic nature of GRBs, inferred by the necessity for the expanding fireball shell to be optically thin, requires bulk Lorentz boost factors in the range $100\lesssim\Gamma\lesssim1000$, with typical values of $\sim300$ for isotropic emission geometry. This has been confirmed\footnote{Lorentz factors greater than 100 have been inferred from the afterglow analyses of GRB 990123 \citep{Soderberg:2002} and GRB 021211 \citep{Kumar:2003}.} via the observation of apparent super-luminal motion in the radio afterglow of GRB 030329 \citep{Taylor:2004}. \emph{Since we are fundamentally limited by the optical opacity of the source at the early stages of its inception, perhaps the next evolution in our understanding of the microphysics of the central engine will be realized via the detection of non-electromagnetic emission such as neutrinos and/or gravitational waves.}

\section{Non-Electromagnetic Emission: A New Window for GRB Progenitors}

The enigma of GRBs, a vestige of their serendipitous discovery over three decades ago \citep{Vela:1973}, is perhaps only rivaled by the phenomenon of cosmic rays, whose sources have remained veiled for nearly a century, by the curvilinear trajectories imposed by intervening magnetic fields (for energies $\lesssim10^{19}$ eV). Ultra high energy cosmic rays (UHECRs\footnote{It is anticipated that protons of energy $\epsilon_{p}\gtrsim10^{19}\:\text{eV}$ are constrained to within a distance of $\lesssim100$ Mpc due to absorption on the cosmic microwave background ($p+\gamma_{\text{CMB}}\rightarrow\pi^{+}+n$), leading to the Greisen-Zatsepin-Kuzmin (GZK) limit ($\sim5\times10^{19}$ eV) \citep{Greisen:1966,ZatsepinKuzmin:1966} and corresponding GZK-neutrino flux \citep{EngelSeckelStanev:2001}, which may help resolve this apparent paradox \citep{SeckelStanev:2005}.}) have been observed with energies of $\sim10^{20}$ eV \citep{Bird:1993}, demonstrating that Nature harnesses astrophysical accelerators that exceed the output of their terrestrial counterparts by several orders of magnitude.

The observed isotropic \citep{Briggs:1996} and cosmological \citep{Metzger:1997} spatial distribution of GRBs, coupled with the fact that the energy injection rates of UHECRs and GRBs are similar, resulted in the phenomenological suggestion that GRBs may be the sources of UHECRs \citep{Waxman:1995,Vietri:1995}. Canonical fireball phenomenology, in the context of hadronic (Fermi) acceleration within the astrophysical jet, predicts a taxonomy of correlated MeV to EeV neutrinos from GRBs of varying flavor and arrival times. Ideal for detection are $\sim$ TeV-PeV muon neutrinos \citep{WaxmanBahcall:1997}, which arise as the leptonic decay products of photomeson interactions ($p+\gamma \rightarrow \Delta ^{+} \rightarrow \pi ^{+} + [n] \rightarrow \mu^{+}+\nu_{\mu} \rightarrow e^{+} + \nu_e +$ $\bar{\nu}_{\mu}+\nu_{\mu}$), within the internal shocks of the relativistic fireball. Since the prompt $\gamma$-rays act as the ambient photon target field, these neutrinos are expected to be in spatial and temporal coincidence, which imposes a constraint that is tantamount to nearly background-free searches \citep{Stamatikos:2005b} in Antarctic Cherenkov telescopes such as AMANDA (Antarctic Muon and Neutrino Detector Array) and IceCube.

AMANDA has produced the most stringent upper limits upon correlated multi-flavored neutrino emission from GRBs \citep{Achterberg:2007,Achterberg:2008}, which are already at the Waxman-Bahcall level. IceCube, AMANDA's km-scale successor, is currently in advanced stages of construction, with over half of its 80 strings deployed, and is already larger than AMANDA. IceCube can cover from $\sim$100 GeV - $\sim$1 PeV for muon neutrinos (which is the direction reconstruction channel, good to a resolution of $\sim1^{\circ}$ over the northern hemisphere, i.e. half sky coverage). All flavor searches, including tau neutrinos, rely on the cascade channel which has all sky coverage, but less resolution, with energies up to $\sim$1 EeV. Both muon and cascade channels are being utilized to search for neutrino emission from GRBs via triggered and un-triggered approaches. This includes efforts toward producing neutrino-based GCN alerts, which are predicated upon rolling temporal searches of neutrino multiplets ($n\gtrsim2$) \citep{KowalskiMohr:2007}. Although the event rate is small ($\sim10^{-3}$ events/GRB/km$^{2}$), fluctuations yielding $\mathcal{O}$(1) event may be realized in exceptional discrete cases such as GRBs 030329 \citep{Stamatikos:2005b} and 080319B \citep{Abassi:2009}, for up to $\sim$10 events per year \citep{Guetta:2004}. The full km$^3$ (Gton) detector is anticipated to be complete by $\sim$2010, and should run for $\sim20$ years. However, since data acquisition began during the construction phase, an accumulated (100 TeV) effective area of $\sim1$ km$^2$ has already been achieved.

Classical GRBs exhibit a bimodal duration distribution at $\sim$2 sec \citep{Kouveliotou:1993}, effectively separating the burst population into short (SGRBs) and long (LGRBs) classes. A connection between some LGRBs and Type Ic supernovae, posited from the death of massive (Wolf-Rayet) stars, has been forged in pairings such as those between GRB 030329 and SN2003dh \citep{Hjorth:2003}, in support of the collapsar progenitor model \citep{Woosley:1993}. This indicates that GRBs signify the birth of black holes, which seems to account for the the association of LGRBs with star forming regions. Meanwhile, the association of GRB 050509B with an elliptical host galaxy \citep{Gehrels:2005} supports the idea that SGRBs may arise from compact binary mergers such as the coalescence of neutron stars ($\sim1.4$ M$_{\odot}$) with other neutron stars (DNS) or black holes (BH-NS/BH-BH) \citep{Bloom:2006}.

The extreme gravitational environments of GRB progenitors are believed to generate gravitational waves (GWs), i.e. vibrations of space-time generated by the acceleration of both mass and energy predicted by Einstein's General Theory of Relativity. Hence, both SGRBs and LGRBs are also likely sources of detectable GWs \citep{Nakar:2006,Kobayashi:2003}, including precursor and/or delayed emission of $\mathcal{O}$(100) sec, which are being hunted across the world via several GW interferometric detectors (IFOs) \citep{Abbott:2008}. Stellar-mass black holes ($\sim10$ M$_{\odot}$) generate high-frequency gravitational waves $(1\lesssim f_{H} \lesssim1000$ Hz) and are detectable by ground-based IFOs such as LIGO. Super-massive black holes (SMBH, $\gtrsim10^{4}$ M$_{\odot}$) produce lower-frequency $(10^{-4}\lesssim f_{L} \lesssim-1$ Hz) signals, which require space-based IFOs such as LISA, to avoid terrestrial seismic noise. Due to NS tidal disruption, BH-NS mergers are expected to have the largest amplitudes within frequency bands $300\lesssim f \lesssim 1000$ Hz, which are ideal for LIGO. While SMBH may be detected in coincidence with GRBs even at cosmological distances \citep{deAraujo:2001}, DNS/NS-BH/BH-BH mergers would require distances of $\sim20/50/100$ Mpc in LIGO, or $\sim300/650/1600$ Mpc for LIGO-II \citep{CutlerThorne:2002}. These distances are within the proximity of SGRBs found by Swift at $\overline{z}\sim0.4$, which corresponds to a comoving radial distance of $\sim1500$ Mpc using $\Lambda_{\text{CDM}}$ cosmological parameters. Thus, GWs detection is possible with LIGO.

\section{Probing the High-z Cosmos with GRBs}

The Swift MIDEX explorer mission, comprised of the wide-field ($\sim1.4$ sr, half-coded) hard X-ray (15-150 keV) Burst Alert Telescope (BAT), and the narrow-field (0.2-10 keV) X-Ray (XRT) and (170-600 nm) Ultraviolet-Optical (UVOT) Telescopes, has revolutionized our understanding of GRBs. The intrinsic multi-wavelength instrumentation, coupled with a rapid ($\lesssim$ 100 seconds) autonomous slew capability, has ushered in an unprecedented era of source localization precision $\left(\lesssim1^{\prime}-4^{\prime}\right)$ that is disseminated in real-time ($\sim10$ seconds) via the GCN, thus spear-heading international ground-based and satellite multi-wavelength follow-up campaigns. Swift's unique dynamic response and spatial localization precision, in conjunction with aforementioned correlative ground-based follow-up efforts, has resulted in redshift determinations for $\sim133$ GRBs, including the most distant cosmological explosion, GRB 080913 at $z=6.695\pm0.025$ \citep{Greiner:2008}, which has begun to constrain progenitor models \citep{Belczynski:2008}. Selection effects, such as detector composition and long accumulation timescales, bias BAT towards long, soft GRBs with lower characteristic photon energy $\left(\text{E}_{\text{peak}}\right)$. Consequently, BAT GRBs comprise a separate statistical class, as is demonstrated by their fluence and redshift distributions $\left(\bar{z}\approx2.3\right)$, from classical Burst and Transient Source Experiment (BATSE) GRBs. High-z GRBs afford an unprecedented opportunity to probe the earliest epoch of stellar formation via the detection of Population I, II (normal) and III (massive) stars. It has been estimated that $\sim10\%$ of Swift GRBs will have $z\gtrsim5$, while $\sim0.5\%$ will have $z\gtrsim10$ (Bromm \& Loeb 2006). This is a reasonable prediction given Swift's observational record, which provides motivation for a tremendous discovery potential. Population III stars, whose first light heralded the end of the cosmic dark ages, are anticipated to have been (i) forged out of metal-free gas at $z\gtrsim10$, (ii) very massive $\left(M_{*}\gg M_{\odot}\right)$ and (iii) instrumental in the thermal and chemical evolution of the intergalactic medium (IGM). Afterglow measurements could reveal high-z GRBs via intergalactic Ly$\alpha$ IR absorption.

With funding recommendations until $\sim2010$ and an estimated orbital lifetime of $\sim15$ years, it is anticipated that the Swift mission will continue to operate until $\sim2017$. Fermi, which is comprised of the ($<$20 MeV to $>$300 GeV) Large Area Telescope (LAT) and the (10 keV - 30 MeV) Gamma-ray Burst Monitor (GBM), launched on June 11, 2008 and has an anticipated mission lifetime of $\sim10$ years, taking it into $\sim2018$. Fermi and Swift, which collectively span $\sim 11$ energy decades, have already commenced correlated GRB observations for over a dozen bursts. It is anticipated that such joint studies will be possible a few times per month \citep{Stamatikos:2008f}. Swift's intrinsic X-ray sensitivity is especially advantageous for high-z GRB detection, since the source frame peak energy would be cosmologically redshifted into a lower detector band pass via $\text{E}_{\text{peak}}^{\text{source}}=\text{E}_{\text{peak}}^{\text{observer}}(1+z)$. This consequence has been exploited in the projected designs of JANUS and EXIST.

JANUS, which will be comprised of the $\sim4$ sr (1-20 keV) X-Ray Flash Monitor (XRFM), the (0.7-1.7 $\mu m$) Near-IR Telescope (NIRT) and the (0.02-1.5 MeV) High-Energy Monitoring Instrument (HEMI), is a contending NASA SMEX mission accepted for phase A, with anticipated launch in $\sim2013$. JANUS will utilize high-z GRBs ($6\lesssim z\lesssim12$) to explore star/galaxy formation and reionization in the early Universe. XRFM's soft energy trigger would enable the detection of XRFs and GRBs at $z\gtrsim6$. JANUS will facilitate the identification of high-z GRBs, which has been difficult to date due to the absence of reliable high-z indicators \citep{Ukwatta:2009}, thus removing the burden of ground-based spectroscopic observations. The NIRT would enable JANUS to provide real-time redshifts (to $\sim5\%$ precision) for $\sim50$ GRBs at $z\gtrsim5$, with $\sim8$ GRBs at $z\gtrsim8$, thus probing deeper into the phase space currently being explored by Swift, with a similar annual GRB rate of $\sim100$.

The EXIST hard X-ray survey, recommended in the 2001 Decadal Survey and created for the Beyond Einstein Program as the Black Hole Finder Probe, is currently in development as ProtoEXIST. EXIST will be comprised of the (5-600 keV) High Energy Telescope (HET), which will localize high-z ($\gtrsim7$) GRBs within $\lesssim100$ sec to an accuracy of $\lesssim10^{\prime\prime}$ at $\sim90\%$, thus allowing for rapid ($\gtrsim1-2$ min) onboard follow-up via the (1.1 m aperture) optical (0.25-1.0 $\mu m$)/NIR (0.9 - 2.5 $\mu m$) telescope (IRT) for imaging and spectroscopy. The wide FOV ($45^{\circ}$ radius at a coding fraction of $\sim25\%$) allows for nearly all sky coverage every 3 hours, which increases the likelihood of triggering on transients such as GRBs. With over an order of magnitude greater area ($\sim5.5$ m$^{2}$) than Swift, EXIST would provide superior sensitivity (complementing Fermi and JANUS) for utilizing GRBs as high-z probes. EXIST's projected GRB rate of $\sim2-3$ day would significantly increase the current annual rates of $\sim200$ and $\sim100$ by Fermi-GBM and Swift-BAT, respectively.

\section{Synthesis \& Broad Science Impact}

GRBs are beacons for multi-messenger astronomy that serve as astrophysical laboratories for Special/General Relativity, particle astrophysics and cosmology. In the electromagnetic regime, Swift's dynamic response and localization precision will complement prompt emission from Fermi, JANUS and EXIST, while facilitating ground-based follow-up via the GCN. Such unprecedented broad-band electromagnetic capability will enable joint GRB data sets that will enhance our understanding of burst parameter classifications, enable routine determinations of E$_{\text{peak}}$ values, explore GRB emission geometry, and help test the viability of various redshift estimation methods \citep{Stamatikos:2008b}. In addition, a more accurate normalization between prompt and afterglow emissions will facilitate the determination of GRB energy budgets, while enabling the investigation of spectral and temporal evolution \citep{Stamatikos:2009} over unprecedented decades of energy. Meanwhile, investigations made by missions such as JANUS and EXIST would utilize GRBs as probes for the high-z Cosmos, via the study of Population III stars and cosmic reionization.

The detection of non-electromagnetic emission from GRBs, via high-energy neutrinos and/or GWs, has the potential for broad scientific breakthroughs. A positive detection of high-energy neutrinos would confirm hadronic acceleration in the relativistic GRB-wind providing critical insight to the associated micro-physics of the fireball while revealing an astrophysical acceleration mechanism for UHECRs, thus resolving a century old enigma. Furthermore, since GRBs are cosmological beacons, the detection of a neutrino beams along astrophysical baselines would help determine intrinsic leptonic properties \citep{Weiler:1994}. The detection of GWs would verify a key prediction of Einstein's General Theory of Relativity, thus turning the universe into an astrophysical laboratory that could  facilitate a solution to problem of quantum gravity. Since both prompt $\gamma$-rays and TeV neutrinos are generated far from the GRB progenitor(s), GWs may be the only viable mechanism of probing the vicinity of the black hole during its genesis. Furthermore, although observational evidence already points towards separate progenitor mechanisms for SGRBs and LGRBs, the detection of GWs would provide definitive proof by exposing compact binary mergers as SGRB progenitors.

High-energy neutrinos and GWs function as unique (albeit elusive) cosmic messengers, which facilitate a new visualization of the Cosmos during its most dynamic processes, while simultaneously complementing the high energy regime of the electromagnetic spectrum. \emph{\textbf{Since neutrinos and GWs are expected to accompany GRBs, observational requirements for correlation studies over the next decade should include:}}

\begin{enumerate}
\item \textbf{\emph{prolific, well-localized GRB triggers with energy spectra and redshift}} - these would help reduce the background, increase the likelihood of detecting extraordinary events, and help model accompanying neutrino/GWs emission from GRBs;
\item \textbf{\emph{a bias towards either SGRBs and/or softer triggering band passes}} - these would minimize the duration ($\lesssim2$ sec, reducing background) and proximity ($z\lll1$, increasing neutrino/GW signal) of GRBs.
\end{enumerate}

Ultimately, multi-messenger astronomy, in the context of GRBs, optimizes discovery potential by enhancing the holistic science return of observatories such as Swift, Fermi, IceCube, LIGO, JANUS, EXIST and LISA, without additional demands upon mutual mission resources, throughout their respective contemporaneous tenure over the next decade. Given such tremendous discovery potential for science synergy and impact, our view of the Cosmos is bound to change forever.

\bibliographystyle{apj}

\bibliography{Astro2010_GRB_Whitepaper}

\end{document}